\documentclass[prl,twocolumn,showpacs,preprintnumbers,amsmath,amssymb,superscriptaddress]{revtex4}
\usepackage{graphicx}% Include figure files
\usepackage{epsfig}% Include figure files
\usepackage{rotating}% Include figure files
\usepackage{dcolumn}% Align table columns on decimal point
\usepackage{bm}% bold math

\newcommand{\lsim}{\raisebox{-0.13cm}{~\shortstack{$<$ \\[-0.07cm] $\sim$}}~} 
\newcommand{\gsim}{\raisebox{-0.13cm}{~\shortstack{$>$ \\[-0.07cm] $\sim$}}~}

\newcommand{\beq}{\begin{eqnarray}} 
\newcommand{\eeq}{\end{eqnarray}} 

\begin{document}

%\preprint{LPT-ORSAY-09/0XX}

\title{The forward--backward asymmetry of top quark production at the  Tevatron
\\ in warped  extra dimensional models}

\author{Abdelhak Djouadi}
\affiliation{Laboratoire de Physique Th\'eorique, Universit\'e Paris XI, 
F-91405 Orsay Cedex, France.}
%\email{djouadi@th.u-psud.fr}
%%%%%%%%%%%%%%%%%%%%%%%%%%%%
\author{Gr\'egory Moreau}
\affiliation{Laboratoire de Physique Th\'eorique, Universit\'e Paris XI, 
F-91405 Orsay Cedex, France.}
%\email{moreau@th.u-psud.fr}
%%%%%%%%%%%%%%%%%%%%%%%%%%%%
\author{Fran\c cois Richard}
\affiliation{Laboratoire de l'Acc\'el\'erateur Lin\'eaire, 
Universit\'e  de Paris--Sud,  F--91898 Orsay Cedex, France.}
%\email{richard@lal.in2p3.fr} 
%%%%%%%%%%%%%%%%%%%%%%%%%%%%
\author{Ritesh K. Singh}
\affiliation{Lehrstuhl f\"ur Theoretische Physik II,  Universit\"at W\"urzburg,  
D--97074 W\"urzburg, Germany.}
%\email{singh@physik.uni-wuerzburg.de}

\begin{abstract} 
The CDF and D0 experiments have  reported on  the measurement of the
forward--backward asymmetry of top quark pair production at the Tevatron and the
result is that it is more than 2 standard deviations above the predicted
value in the Standard Model. This has to be added to the longstanding anomaly in
the  forward--backward asymmetry for bottom quark production at LEP which is 3
standard deviations different  from the Standard Model  value. 
The discrepancy in the bottom asymmetry can be accounted for by the contributions of
Kaluza--Klein excitations of electroweak gauge bosons at  LEP
in warped extra dimensional models in which the 
fermions are localized differently along the extra  dimension so that the gauge 
interactions of  heavy third generation fermions are naturally different from
that of light fermions. In this paper, we show that it is more difficult to elaborate a model
generating a significant 
top asymmetry in a similar way - through exchanges of Kaluza--Klein gluons at the Tevatron - 
due to the indirect constraints originating from precision electroweak data.

\end{abstract} 

\maketitle

Apparently, something is indeed  rotten in the kingdom of third generation
quarks.  Adding to the longstanding anomaly of the forward--backward (FB)
asymmetry for $b$--quark jets $A_{\rm FB}^b$  measured in $Z$ boson decays at
LEP \cite{PDG,LEPex}, which differs by 3 standard deviations from the Standard
Model (SM) value \cite{AFB-SM}, the CDF and D0 collaborations have reported
results  \cite{CDF,Tevatron} on the measurement of the FB asymmetry of top quark
pairs produced at the Tevatron,  $A_{\rm FB}^t$,  that are not consistent with
the SM expectation. In particular, the latest and most precise result from the
CDF collaboration  \cite{CDF}, using 3.2 fb$^{-1}$ data,  gives for this
asymmetry  in the $p\bar p$ laboratory frame
\beq 
\label{Afbex}
A_{\rm FB}^t=0.19 \pm 0.065 \, ({\rm stat}) \pm 0.024 \, ({\rm syst})\, .
\eeq

In the SM, this asymmetry is predicted to be vanishing at first order in QCD. 
Indeed, a very nice feature of the Tevatron is that it is almost a $q\bar q$
collider for top quark pair production as the process  occurs mainly through
virtual gluon exchange in $q\bar q$ annihilation, with  only a small
contribution from the initiated gluon-gluon fusion channel.  As gluons have only
vector--like couplings to quarks,  the process  does not generate an asymmetry
between quarks and antiquarks and thus, $A_{\rm FB}^t$ is identically
zero~\cite{Foot1}.  The asymmetry is then generated  at  next--to--leading order
(NLO) in QCD by diagrams involving an extra gluon radiation and
(anti)quark--gluon annihilation as well as  from the interference between the
Born gluon  exchange with one--loop box diagrams. These NLO contributions lead
to the  expected value in the SM of \cite{Kuhn} 
\beq  
A_{\rm FB}^t=0.05 \pm 0.015\, .  
\eeq 
In the absence of large higher order contributions~\cite{Stefan}, this leads to
a 2 standard deviation between the experimentally measured and the theoretically
predicted  values. This is in contrast to the total $p\bar p \to t\bar t$
production cross  section at the Tevatron which is measured to be \cite{Review}
\beq
\sigma(t\bar t)^{\rm ex}= 7.0 \pm 0.63 ~{\rm pb} \, ,
\eeq
in a good agreement  with the SM expectation \cite{Mateo,newPDF}, 
\beq
\sigma(t\bar t)^{\rm th}=  7.0^{+0.71}_{-0.79}  ~{\rm pb}\, .   
\eeq

As in the case of the LEP $A_{\rm FB}^b$ anomaly (see Ref.~\cite{dmr} and
references therein), it is very difficult to explain this discrepancy, without
affecting  significantly the well behaved $t\bar t$ cross section, in well
motivated extensions of the SM such as supersymmetric models for
instance~\cite{Foot2}.  Among the very few  attempts that have been made,
examples are the exchange of TeV mass axi-gluons \cite{Kuhn}, colored gauge
bosons which have axial--vectorial couplings to quarks; another possibility
discussed in Ref.~\cite{Rohini}, would be flavor universal colorons  which
occur in gauge group models with an extended color such as topcolor or topcolor
assisted technicolor models. These extensions do note cure the LEP $A_{\rm
FB}^b$ anomaly, though.

In Ref.~\cite{dmr} it has been shown that the discrepancy between the LEP
measured value of $A_{\rm FB}^b$ and the theoretical prediction can be resolved
in the context of variants of the Randall--Sundrum (RS) extra--dimensional model
\cite{RS} in which the SM fermion and bosonic fields are propagating in the
bulk, except for the Higgs boson that is confined on the so--called TeV--brane.
This allows a new interpretation of SM fermion mass hierarchies, if these
fermions are localized differently along  the extra dimension depending on their
nature.  One can then naturally obtain ElectroWeak (EW) interactions for the heavy
third generation fermions that are different from the ones of the light
fermions. More precisely, the $Z$ boson will mix with its Kaluza--Klein  (KK)
excitations and only its overall couplings to third generation fermions are
significantly altered, due to the higher KK gauge boson coupling to the heavy
flavors. An adequate choice of the $b$--quark localization allows to explain the
3$\sigma$ deviation of $A_{\rm FB}^b$,  while keeping all other precision
measurements unaltered. 

In this letter, we show how the same warped extra dimensional scenario that
resolves the LEP $A_{\rm FB}^b$ anomaly could {\it in principle} also 
soften the discrepancy between the measured value of $A_{\rm FB}^t$  at the
Tevatron and its theoretical value. Here again, the apparent $A_{\rm FB}^t$
anomaly could be addressed thanks to the naturally larger KK gauge boson couplings to
the third generation quarks but when trying to construct a realistic scenario
respecting all the precision electroweak measurements we find it more tricky
to significantly enhance $A_{\rm FB}^t$. Nevertheless, the realistic scenario
obtained, where $A_{\rm FB}^t$ is relatively enhanced, 
provides interesting possibilities for the direct search of KK modes.

%%%%%%%%%%%%%%%%%%%%%%%%%%%%%%%%%%%%%%%%%%%%%%%%%%%%%%%%%%%%%%%%%%%%%%%%%%%%%%

The RS  warped extra--dimensional scenario \cite{RS} was originally proposed as
a solution to the gauge hierarchy problem.  It consists of a five--dimensional
theory where the warped extra dimension  is compactified over a $S^{1}/\mathbb{
Z}_{2}$ orbifold.  The fermions possess five--dimensional masses, quantified by
the parameters $c_f$  associated to each multiplet. These various masses
determine the fermion localizations along the  extra dimension. A possible way
to avoid large deviations from the KK states in the set of high precision
EW observables  \cite{PDG,LEPex}, while keeping the mass of the first
KK weak gauge boson excitations  (that  are nearly equal to  the KK gluon mass
$M_{KK}$) as low as the TeV scale, is to extend the SM group by gauging the
custodial symmetry ${\rm SU(2)_L\! \times\! SU(2)_R\! \times\! U(1)_X}$  in the
bulk \cite{ADMS}. In particular, an additional KK $Z'$ boson then arises with a
coupling constant $g_{Z'}$ that is related to the  mixing angle between the $Z$
and $Z'$ bosons.  

In the SM sector of the gauge bosons and light fermions $f \neq b, t$, if the fermion
localization and hence the $c_f$ parameters are such that $c_{\rm light} \gsim
0.5$ \cite{RSloc}, they lead to an acceptable fit of EW data provided 
that $M_{KK} \approx 3$ TeV in the case of the bulk custodial symmetry
\cite{ADMS}. In contrast, for third generation  $Q=t$ and $b$ quarks, the
parameters for right-- and left--handed states  $c_{t_R}$, $c_{b_R}$ and
$c_{Q_L}=c_{b_L} =c_{t_L}$ (as a result of SU(2) symmetry), should be chosen
smaller,   $c_Q \lsim 0.5$, in order to produce relatively large quark masses
\cite{RSloc}. Thus, the corrections to the crucial observables of the heavy
$b$--quark sector at LEP, namely  $A_{\rm FB}^b$ and the partial $Z$ boson decay
width  $\Gamma(Z \to b\bar b)$, and the  Tevatron observables in top quark
production,  $A_{\rm FB}^t$  and $\sigma(t \bar  t)$, have to be treated
separately.

We consider the scenario consisting of the quark multiplets given in \cite{FootNEW}, 
which correspond to our previous choice of representation \cite{dmr} (RSb model). 
For the parameter values, we have found interesting to take $M_{KK} \simeq  2.75$ TeV and:   
\beq  
\label{RSparametersBIS} 
c_{Q_L}=0.35,  c_{b_R}=0.49;   c_{\rm light} \gsim
0.5; \ g_{Z'}=3.1 
\eeq  
which leads to a good fit of the observables in the $b$--quark sector at
LEP (namely $R_b=\Gamma(Z \to b\bar b)/\Gamma(Z \to {\rm hadrons})$ 
and $A_{\rm FB}^b(\sqrt{s})$ at the different energies): a $\chi^2 \simeq 16$  compared to  $\chi^2 \simeq 21$ in the SM
\cite{dmr,AFBRST}, and  a bottom quark mass $m_b \simeq 4$ GeV, which is
acceptable keeping in mind that a full three--flavor treatment (beyond our scope here and let for a future study) 
would even improve the value. Moreover, for the {\it non bottom-top} quark EW observables, it allows 
to obtain  a global fit  that is better than in the SM as shown in \cite{AFBRST}.

Generically in the RS model, the pair production of top quarks in $q\bar q$ 
annihilation does not proceed through gluon exchange (and gluon--gluon fusion) 
only, but also via the exchange of the KK gluon.  The couplings of the first KK
excitation of the gluon  to left-- and right--handed $q \equiv u,d$ quarks are
different and proportional to  $g_S Q(c_{q_{L/R}})$ where $g_S$ is the usual QCD
coupling  and the charges $Q(c_{q_{L/R}})$ are the geometrical factors giving
the ratio to the four--dimensional  effective coupling of the gluon to
$q_{L/R}$; for a light quark $q$ with $c_q \gsim 0.5$ one  has $Q(c_q) \approx
-0.2$, while for the heavy third generation $t,b$ quarks,  $Q(c_{t,b})$ can be
taken close to or larger than  unity. Thus, the KK gluon coupling to quarks is
not vectorial anymore, but has also an axial--vectorial component,
$v_q/a_q\propto Q(c_{q_R})\pm Q(c_{q_L})$. It is this axial--vector component of
the KK gluon coupling which  will generate a FB asymmetry for top quark pair
production at the tree--level.  The angular distribution of the subprocess
$q\bar q \to  t\bar t$ is then given by
\beq
\label{dsigma}
&&\frac{ {\rm d} \hat \sigma }{ {\rm d} \cos \theta_t^*} \propto
2-\beta_t^2\sin^2\theta^*  + \hat s^2 |{\cal D}|^2
\Big[ 8 v_qv_t a_qa_t \beta_t \cos\theta^*  \nonumber \\
&&+ (a_q^2+v_q^2) \left( v_t^2 (2 - \beta_t^2\sin^2\theta^*) + a_t^2
\beta_t^2 (1+\cos^2\theta^*)\right) \Big] \nonumber \\
&&+ \hat 4s {\rm Re} ({\cal D}) \big[
v_q v_t \left( 1- \frac12 \beta_t^2 \sin^2\theta^* \right)
+  a_q a_t \beta_t \cos\theta^* \big]
\eeq
where $\hat s$ is the effective c.m. energy of the subprocess, $\theta^*$  the
scattering angle in the $q\bar q$ frame, $\beta_t = \sqrt{1- 4m_t^ 2/\hat s}$ is
the velocity of the top quark and ${\cal D}=(\hat s -M_{KK}^2  + i\Gamma_{KK}
M_{KK})^{-1}$ the propagator of the KK gluon with mass  $M_{KK}$ and  total
width $\Gamma_{KK}$. To obtain the $p\bar p$ hadronic cross section $\sigma$,
one must then integrate over the angle $\theta^*$, sum over all  contributing
initial quarks and convolute with their parton distribution  functions.  

The FB asymmetry of the top quark is then defined as 
\beq 
A_{\rm FB}^t = \frac{\sigma(\cos\theta_t>0)-\sigma(\cos\theta_t<0)}{\sigma(
\cos\theta_t>0)+\sigma(\cos\theta_t<0)}. 
\eeq 
where now $\theta_t$ is the angle between the reconstructed top quark momentum
relative to the proton beam direction. It is proportional to the factor
in front of  $\cos\theta^*$ in eq.~(\ref{dsigma}), 
\begin{eqnarray}
A_{\rm FB}^t \propto a_qa_t\beta_t \ \hat{s} \ |{\cal D}|^2
\left[ (\hat s - M_{KK}^2) + 4v_qv_t \ \hat s \right],
\end{eqnarray}
which originates from the interference between the gluon and the KK gluon
contributions and also from the pure KK gluon diagram. Note that  $A_{\rm FB}^t$
is non--zero only if both  axial--vector couplings of $g_{KK}$, $a_q$ and $a_t$
are non--zero. The product $a_q a_t$ should be negative  along with  $\hat s <
M_{KK}^2/(1+2v_qv_t)$,  to have a positive $A_{\rm FB}^t$ below the $g_{KK}$
resonance, as is the case with $M_{KK} \approx 3$ TeV.  Thus, one needs to
maximize $a_q a_t$ while keeping  $v_qv_t$ reasonable to achieve a large 
asymmetry. However, if $a_qa_t$ is too large, it will significantly alter
$\sigma(t\bar t)$ which is in accord with the SM. A judicious choice of the 
couplings  $a_q$ and $a_t$ of the $g_{KK}$ excitation is thus required. 

Note that at NLO, there are additional contributions to $A_{\rm FB}^t$, e.g.  
stemming from the interference of the diagram with KK gluon exchange and the  
SM box diagrams; these small corrections will not be considered here. 

%%%%%%%%%%%%%%%%%%%%%%%%%%%%%%%%%%%%%%%%%%%%%%%%%%%%%%%%%%%%%%%%%%%%%%%%%

\begin{figure}[!h]
\begin{center}
\vspace*{-2mm}
\epsfig{file= 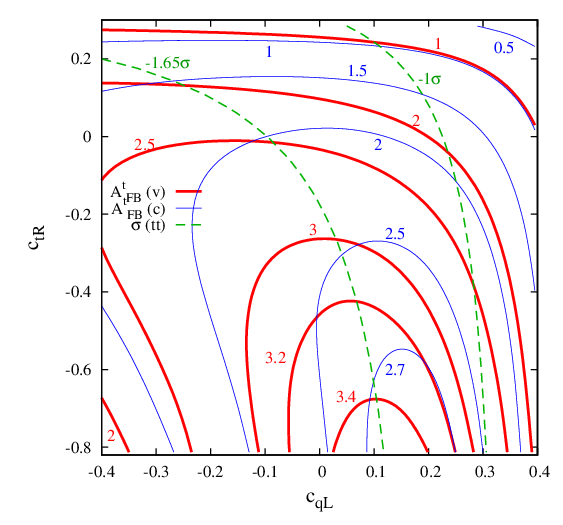,width=9.cm,height=7.cm} 
\end{center}
\vspace*{-.7cm}
\caption{Contour levels in the plane $[c_{q_L}, c_{t_R}]$ ($c_{q_L}$ is for first generation) for the total cross 
section $\sigma(t\bar t)$ (dashed lines) and the forward--backward asymmetry 
$A^t_{FB}$ [in \%] for constant (thin solid lines) and variable (thick solid 
lines) total decay width for the  KK gluon; the other parameters are as  in
eq.~(\ref{RSparametersBIS}).}  
\label{plot:AFBtbis}
\vspace*{-.1cm}
\end{figure}

The numerical results that we obtain are summarized in 
Fig.~\ref{plot:AFBtbis} which displays the contour levels in the plane  $[c_{q_L},
c_{t_R}]$ corresponding,  typically, to the maximum $A_{\rm FB}^t$ asymmetry as
well as the associated $\sigma(t \bar t)$ values (within $\approx 1.65\sigma$ 
of the experimental value i.e. at the 90\% C.L.). 
Note that the domain $c_{t_R}<-0.6$ corresponds typically to too light custodians.
In the figure, we have fixed
the $c$ values of the  right--handed first generation quarks to $c_{u_R} \approx c_{d_R} 
\approx 0.8$.  The chosen range for $c_{q_L}$ with much smaller values than
$c_{u_R}\approx c_{d_R}$, allows substantial parity violating couplings of first
generation quarks to the KK gluon and, hence, a sizable  $A_{\rm FB}^t$.  
It seems that this choice $c_{q_L}<0.5$ requires a certain flavor structure 
among 5D Yukawa couplings for reproducing quark mixing angles.
A study dedicated to the complete three-flavor structure will be performed in \cite{progress}. 

The decrease of $\sigma(t \bar t)$ with both $c_{q_L}$ and $c_{t_R}$ is caused
by the increase of the $g_{KK}$ couplings to $\bar q_L q_L$ and $\bar t_R t_R$
states which dominantly enhance the total $g_{KK}$ decay width. The increase of
$A_{\rm FB}^t$ with the decrease of $c_{t_R}$, which amplifies the difference
between $c_{t_R}$ and the fixed $c_{Q_L}$, finds its origin in the larger parity
violation effect on the four--dimensional $g_{KK} \bar t t$ coupling.  $A_{\rm
FB}^t$ can thus reach sizable values in regions where $\sigma(t \bar t)$ has
values consistent with Tevatron data at the $1.65\sigma$ level.

In the region $c_{t_R} \simeq - 0.5$, $c_{q_L} \sim 0.4$, 
the order of magnitude obtained for top quark mass, $m_t
\approx 10^2$ GeV, is acceptable while, despite of the atypically large
$Q(c_{q_L})$ values for first generation quarks (we had also to take
$c_{s_R} \simeq 0.47 <0.5$), the $Z \bar qq$ 
couplings are in good agreement with all precision data: the measurements at LEP \cite{LEPex},
the NuTeV results \cite{PDG,NuTeV} as well as with the less accurate 
Tevatron and HERA data on quarks $u/d$ \cite{HERA-CDF}. 
This separate analysis of precision constraints on the first two generation quarks will
be described in details in \cite{progress}; the reasons why these constraints can be respected
are the relatively low $c_{q_L}$ value and the existence of cancellations. For exemple,
the small deviations to the $Z$ hadronic width result from 
an approximative cancellation of the $Z'$-induced corrections between $u_L$ and $d_L$ due to isospins
\cite{FOOTroleZp}. 
The deviations to this cancellation are compensated 
by the corrections induced by the KK $Z$ bosons 
and the other quark chiralities/generations (controled by other parameters). 
Interestingly, the NuTeV discrepancy 
on $g_L^2$ for first generation quarks at $1.9 \sigma$ in the SM decreases here down to
$0.7 \sigma$, while we obtain
$g_R^2$ at $0.6 \sigma$ only [weaker experimental accuracy]. 
This significant change w.r.t. SM
is allowed by the absence of correction compensation among different quark chiralities/generations, 
in contrast with the case of the $Z$ hadronic width.

One must remark that for our parameters the total decay width for the KK
gluon turns out to be quite large,  $\Gamma_{KK} \approx 30\% M_{KK}$.   For
such a broad resonance, at least the energy dependent width, if not the full set
of radiative corrections to the $p\bar p \to t\bar t$ process, should be used, 
in much the same way as for the $\rho$ vector--meson exchange in $e^+ e^- \to
\pi^+ \pi^-$ \cite{Gounaris}. We have checked that by doing  so, the
contribution to $A_{\rm FB}^t$ from KK gluon exchange is increased
as illustrated on the figure [see the thick solid contour lines].

In the acceptable region considered above ($c_{t_R} \simeq - 0.5$, $c_{q_L} \sim 0.4$), 
the contribution from the KK gluon exchange is $A_{\rm FB}^t \simeq 2\%$ 
as shown by the contours of Fig.~1 \cite{Foot5}. Adding this contribution to the SM NLO 
value,  one obtains a total of  $A_{\rm FB}^t|^{\rm RS+SM} \approx 7\%$, which
represents a relative improvement over the SM results. The significant gap
remaining beetwen the experimental $A_{\rm FB}^t$ value 
and its theoretical prediction could be explained by the still uncalculated QCD corrections
at NNLO -- possibly important. 
Note finally that there is no fine--tuning of parameters as in a large range of $c$ values, a sizable
$A_{\rm FB}^t$ is obtained.

In order to illustrate the importance of the indirect precision electroweak constraints,
we just mention here an example of different choice of quark representations
given in \cite{Foot3} which could lead to a more important $A_{\rm FB}^t$ increase but 
in which it seems impossible to get realistic values simultaneously for the 
observables on $Z \bar qq$ vertex coming from the atomic parity violation 
and the quark asymmetry $Q_{FB}$ \cite{PDG}. Here, the effects
of KK quark mixing in the first generation sector should be studied in more details.

At this stage, a few important remarks are in order. 

$i)$ Another puzzling feature of the CDF data is that the differential 
cross section with respect to the $p\bar p \to t\bar t$  invariant mass is,
except for the two extreme  points, systematically below the SM expectation
\cite{CDFexcess}. The negative
interference between the contributions of the gluon and its KK excitation can lower
the mass distribution as to fit nicely the data. We expect a tiny excess of
events, ${\cal O}(20\%)$,  for invariant $t \bar t$ masses above 800 GeV; this
excess is found to be more significant in other scenarios, e.g. for slightly
lower $M_{KK}$ values.

$ii)$ The contribution to $A^t_{\rm FB}$ and $\sigma(t \bar t)$ from the
exchange of the KK excitations of EW gauge bosons are not dominant for
the model considered in the present paper.

$iii)$  At the Large Hadron Collider, gluon KK states could be slightly more
produced in our scenario where the KK gluon coupling to first generation quarks 
is enhanced (as $c_{q_L}<0.5$, in contrast with usual assumptions, 
which pushes the quark wave functions towards the KK profile) 
and top charge asymmetry could be measured \cite{KKgluonLHC},
confirming or invalidating  the present scenario; all this might be, however, rather
challenging. 

%%%%%%%%%%%%%%%%%%%%%%%%%%%%%%%%%%%%%%%%%%%%%%%%%%%%%%%%%%%%%%%%%%%%%%%%%%%%%

In conclusion, we have proposed an RS scenario in which the theoretical value 
of $A_{\rm FB}^t$ is relatively closer to the value measured at the Tevatron, compared to the SM situation. 

If the deviation persists in data and the relevant higher order corrections to $A_{\rm FB}^t$ in the SM
can mainly explain the experimental result, then the warped extra dimensions
could be play a partial role in the interpretation of this discrepancy.

\smallskip

\noindent \textbf{Acknowledgments:} The work of AD and GM is supported by
HEPTOOLS while the work of RKS is supported by the German BMBF under contract
05HT6WWA. We thank Z.~Zhang for discussions.


\begin{thebibliography}{999}
\bibitem{Foot1} There are also contributions to the $q\bar q\to t\bar t$
process  from the photon and the $Z$ boson.  However, because of the small
EW coupling compared to the QCD coupling  and, since these diagrams do
not interfere with the one with gluon exchange as a result of color
conservation, the contributions are extremely small, ${\protect \lsim 0.15\%}$.
\bibitem{Foot2} We have evaluated the contribution of the box diagrams involving
squarks and gluinos in the minimal supersymmetric extension of the SM (MSSM) to
$A_{\rm FB}^t$ and found that, even for squark and  gluino masses close to 100
GeV, the interference with the tree SM diagram  with gluon exchange cannot
exceed the level of $10^{-4}$.
\bibitem{FOOTroleZp} Note the new motivation of the custodial symmetry
bringing a $Z'$ boson which plays a central role in EW fits for the first/third quark generations.
\bibitem{Foot5} We recall again that these values are given in the $p\bar p$
rest frame;  in the $t\bar t$ partonic frame, they have to be multiplied by a
factor $\approx 1.5$ so that the maximal asymmetry obtained in this case is
$A_{\rm FB}^t \approx 3\%$.
\end{thebibliography}
\end{document}